\documentclass[12pt]{article}
\pdfoutput=1
\usepackage{jheppub}
\usepackage[utf8]{inputenc}
\usepackage{verbatim}
\usepackage{amsmath}
\usepackage{amssymb}
\usepackage{amsthm}
\usepackage{slashed}
\usepackage{amsfonts}
\usepackage{hyperref}

\newcommand{\be}{\begin{equation}}
\newcommand{\ee}{\end{equation}}
\newcommand{\bfig}{\begin{figure}\begin{center}}
\newcommand{\efig}{\end{center}\end{figure}}
\newcommand{\vn}{\vec{\nabla}}
\newcommand{\lan}{\langle}
\newcommand{\ran}{\rangle}
\newcommand{\wt}{\widetilde}
\newcommand{\Tr}{\mathrm{Tr}}
\newcommand{\bi}{\begin{itemize}}
\newcommand{\ei}{\end{itemize}}

\begin{document}
\title{Black holes in quantum gravity}
\author{Daniel Harlow}
\affiliation{Center for Theoretical Physics\\ Massachusetts Institute of Technology, Cambridge, MA 02139, USA}
\emailAdd{harlow@mit.edu}
\abstract{This chapter gives an overview of the quantum aspects of black holes, focusing on the black hole information problem, the counting of black hole entropy in string theory, and the emergence of spacetime in holography.  It is aimed at a broad physics audience, and does not presuppose knowledge of string theory or holography.}
\maketitle
\section{Introduction}
The topic of this chapter is the quantum mechanics of black holes.  It must be stated at the outset that this is not an experimental subject, and it does not seem likely to become one in the near future.  Hawking has predicted that black holes radiate quanta whose wavelength is comparable to the size of the black hole, at a rate of roughly one quanta per light-crossing time.  For the smallest astrophysical black holes with $M\sim M_{\odot}$ this amounts to producing of order $10^5$ photons per second, each of which has a wavelength of order a kilometer.  Detecting this flux would be quite challenging even in the vicinity of such a black hole, and in fact the closest known black holes are kiloparsecs away.  Moreover detecting this flux, while important, would only be the initial challenge.  To truly get at the deep conceptual puzzles underlying black hole physics, we likely would need to measure an $O(1)$ fraction of the Hawking radiation.  One obvious problem is that the time it would take a solar mass black hole to evaporate is of order $10^{63}$ years, which is a long time to gather data.  Another problem is that we would likely lose a sizable fraction of that radiation since it would consist of gravitons, which are far more difficult to detect than photons.  And even worse, if we indeed lose the gravitons then to really settle things we would most likely need to do a quantum computation on the collected radiation whose runtime is exponential in the black hole entropy, which would take a further time of order $10^{10^{80}}$ years.  To avoid all these problems we would likely need to create our own black holes, which we would want to be much smaller than the black holes which nature creates for us (a possible exception to this statement would be if small ``primordial'' black holes were created by density fluctuations in the early universe, but so far we have not seen any).  This too is difficult: astrophysical black holes are created by gravitational collapse, but for less massive objects the strength of the other forces relative to gravity increases and gravitational collapse becomes difficult to achieve (this is why there is a lower bound on the mass of astrophysical black holes).  Creating ``small'' black holes would require us to bring their constituents close enough that they would be within their mutual Schwarzschild radius, and if the number of constituents is $O(1)$ this requires collision energies of order the Planck mass $m_p\sim 10^{-8} \,\mathrm{kg}$, which exceeds the collision energy of the Large Hadron Collider by a factor of order $10^{15}$. 

Given all these obstacles, why study the quantum mechanics of black holes?  There are various answers which have been given, but perhaps the most compelling is the following: we know that black holes exist, and we know that our world is quantum mechanical, and so it must be that nature finds some way to combine them.  And moreover, as we will see below, this combination is surprisingly difficult: the naive way of doing it leads to apparent contradictions, which it seems cannot be avoided without a serious modification of the laws of physics as we currently understand them.  This is similar to the situation Einstein found himself in in 1905: he knew that Newtonian mechanics and Maxwell's electromagnetic theory were both on solid experimental footing, but he realized that they could not be consistently applied to objects moving close to the speed of light.  In understanding this incompatibility it was very convenient for Einstein to consider thought-experiments involving trains moving close to the speed of light, even though it was (and is) very unlikely that such trains could be constructed.  Indeed thinking about them led Einstein to a new theory, relativistic mechanics, which removed the contradictions but reduced to the previous two theories in the appropriate limits.  Similarly for quantum black holes, the hope is that in resolving the apparent contradictions we will be led to a self-consistent theory of quantum gravity.  And if we are lucky, as Einstein was, that theory will make predictions about other situations besides the one where the contradictions are most clear, and those predictions may well be testable.  This is not a baseless hope: we already know that the calculation of Hawking radiation is closely related to the calculation of the spectrum of density perturbations produced during cosmological inflation, which have already been observed in the fluctuations of the cosmic microwave background, and it is quite plausible that a deeper understanding of quantum gravity could lead to testable predictions for cosmology.  

There has also been another benefit to thinking about the quantum mechanics of black holes: many of the natural questions have turned out to have broader relevance in theoretical physics and computer science, leading to new ideas in fields such as quantum chaos, superconductivity, quantum cryptography, complexity theory, relativistic plasma physics, particle physics, and even mathematics.  Conversely ideas from those fields have been useful in better understanding black holes, and these days most people working on the subject are fluent in a broad range of topics across theoretical physics.  There is even a growing dialogue with atomic physics experimentalists, who may be able to simulate some simple quantum gravity systems in the laboratory and thus test some of their predictions directly.\footnote{Despite some claims to the contrary, such experiments cannot be viewed as genuine tests of quantum gravity: the true theory of quantum gravity in our world need not be the one we choose to simulate.  These experiments are better understood as a way to ``numerically'' illustrate consequences of possible quantum gravity models which are difficult to study analytically, just as is done e.g. in the study of turbulence.}

The remainder of this chapter will review the quantum mechanics of black holes, beginning with the seminal work of Bekenstein and Hawking in the 1970s, which was based on a semiclassical picture of quantum gravity.  We will see that to proceed further some assumptions need to be made about the broader question of how to combine quantum mechanics and gravity.  Various approaches to this problem have been proposed, and so far none of them have led to a theory of quantum gravity which is both well-defined and realistic.  On the other hand by far the most successful approach has been that based on string theory and holography, and so that is the approach we will mostly adopt.\footnote{Other approaches include ``loop quantum gravity'' and ``causal dynamical triangulation'', which try to explicitly discretize spacetime at short distances, but so far these approaches have not been able to produce a Lorentz-invariant theory at low energies in $3+1$ dimensions. They also have not been able to account for the thermodynamics of black holes that we review below.  String theory and holography have solved both problems.  Another approach is ``asymptotic safety'', which postulates the existence of a strongly-coupled conformal phase of gravity at short distances, but so far there is no evidence for such a phase and this approach also seems unlikely to reproduce black hole thermodynamics.}    

\section{Classical black holes}
We begin with the classical theory of black holes in $3+1$ dimensions.\footnote{Readers who are unfamiliar with the material reviewed here may wish to consult \cite{Harlow:2014yka} for a more gentle introduction.  Or any textbook on general relativity, for example \cite{carroll2019spacetime}.}  Realistic black holes are formed from matter collapse, but this is somewhat messy so it is simpler to first consider the idealized \textit{Schwarzschild geometry}, which in one set of coordinates has metric
\be\label{schdef}
ds^2=-f(r)dt^2+\frac{dr^2}{f(r)} +r^2d\Omega_2^2
\ee
with
\be
f(r)\equiv 1-\frac{2GM}{r}.
\ee
Here $d\Omega_2^2=d\theta^2+\sin^2\theta d\phi^2$ is the round metric on a unit $\mathbb{S}^2$, the radial coordinate $r$ runs from zero to $\infty$, and we are measuring time in light-meters so $c=1$.  The function $f(r)$ approaches one at large $r$, so this geometry resembles Minkowski space there.  The metric appears singular at two different values of $r$, $r=0$ and $r=r_s\equiv 2GM$, but the nature of these singularities is quite different: the singularity at $r=r_s$, which is called the \textit{horizon}, is an artifact of a bad choice of coordinates, while the singularity at $r=0$ is a genuine singularity which crushes anyone who approaches it.  The horizon does have \textit{global} significance however: for $r<r_s$ we have $f<0$, so the coordinate $r$ becomes timelike and $t$ becomes spacelike.  In particular this implies that anyone who enters the region with $r<r_s$ is doomed to reach the true singularity at $r=0$ and be destroyed: someone with $r<r_s$ cannot stop $r$ from decreasing anymore than you can stop $t$ from increasing out near $r=\infty$.

\bfig
\includegraphics[height=5cm]{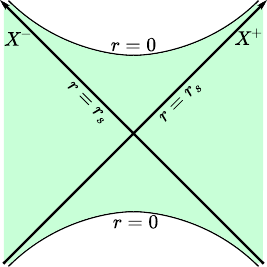}
\caption{The Schwarzschild geometry in Kruskal coordinates.  Light cannot propagate from one exterior to another, so it describes a ``non-traversable wormhole''.}\label{kruskalfig}
\efig
To get a better global picture of the Schwarzschild geometry we need to change coordinates.  One good choice is \textit{Kruskal coordinates}, which are defined by 
\begin{align}\label{Kruskaldef}
X^\pm \equiv \pm r_s e^{\frac{r_*\pm t}{2r_s}},
\end{align}
with
\be
r_*\equiv r+r_s\log\left(\frac{r-r_s}{r_s}\right),
\ee
and in terms of which the metric is
\be
ds^2=-\frac{2r_s}{r}e^{-r/r_s}\left(dX^+dX^-+dX^-dX^+\right)+r^2d\Omega_2^2.
\ee
Here the relationship between $X^\pm$ and $r$ is
\be
X^+X^-=(r_s-r)r_s e^{r/r_s}.
\ee
The basic idea of Kruskal coordinates is that outgoing/ingoing radial null geodesics are curves of constant $X^+$/$X^-$.  The full geometry is shown in figure \ref{kruskalfig}, note that there are \textit{two} regions where $r\to\infty$, and they are connected by a ``wormhole'' with both ``future'' and ``past'' singularities where $r=0$.  

\bfig
\includegraphics[height=4cm]{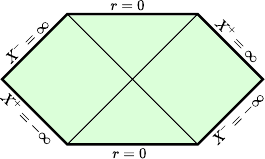}
\caption{The Penrose diagram for the Schwarzschild geometry.  Time goes up, the diagonal lines are the horizons, and light moves on $45^\circ$ lines.  The horizons split the spacetime into four regions, which are sometimes called the future/past interiors and the left/right exteriors, and the diagram makes it clear that there can be no causal communication between the two exteriors.  It is also clear that any causal observer who enters the future interior will not leave it without meeting the future singularity where $r=0$.}\label{penroseschfig}
\efig
In practice it is often useful do an additional coordinate transformation, for example $y^\pm \equiv \arctan(X^\pm)$, which brings in infinity to a finite distance while preserving the feature of Kruskal coordinates that radial null geodesics are lines of constant $y^\pm$.  This allows us to represent the full geometry including infinity in a compact diagram called a Penrose diagram:   the Penrose diagram for the Schwarzschild geometry is shown in figure \ref{penroseschfig}.  

\bfig
\includegraphics[height=5cm]{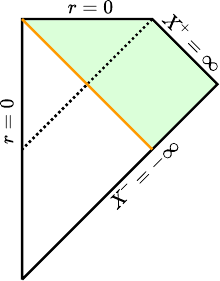}
\caption{The Penrose diagram for a black hole formed by a collapsing shell of photons.  The horizon is the dashed black line, while the photon shell is the solid orange line.  The shaded green region is a piece of the Schwarzschild geometry, but there is now only one asymptotic region.}\label{penrosecollapsefig}
\efig
Real astrophysical black holes that form from matter collapse are not described by the Schwarzschild geometry.  For one thing they do not involve a second asymptotic universe, at least as far as we know!  On the other hand we expect the geometry to the future of the infalling matter to be described by a piece of the Schwarzschild geometry to an excellent approximation.  Indeed we can simplify life by taking the collapsing matter to consist of a spherical shell of photons moving radially inward, in which case the geometry to the future of the shell is \textit{exactly} a piece of the Schwarzschild geomtetry, while the geometry to the past of the shell is exactly that of empty flat space.  The Penrose diagram for a black hole formed in this way is shown in figure \ref{penrosecollapsefig}.  The horizon is now defined as the boundary of the set of points which cannot send signals out to large $r$, and in figure \ref{penrosecollapsefig} it is shown as a dashed line.  In particular note that the horizon extends down into the region before the shell has arrived: we ourselves could already be behind the horizon of a large black hole which has not yet formed.

\section{Quantum fields near black holes}
Much more could be said about the classical theory of black holes, but we now turn to quantum effects.  The simplest quantum effects near black holes are those which arise from the quantum properties of ordinary matter fields in the vicinity of the black hole.  The full standard model of particle physics has many matter fields of various spin: the vector gauge bosons mediating the strong and electroweak forces, the spinors describing quarks and leptons, and the scalar Higgs field.  For the purposes of this article these are all qualitatively similar, so it is convenient to study a simpler theory where the only matter field is a single real massless scalar $\phi(x)$ which has no non-gravitational interactions.  The Lagrangian density of this theory coupled to Einstein gravity is
\be
\mathcal{L}=\frac{R}{16\pi G}-\frac{1}{2}\partial_\mu \phi \partial_\nu \phi g^{\mu\nu},
\ee
where $G$ is Newton's constant and $R$ is the Ricci scalar.  This theory has two dynamical fields, the scalar $\phi$ and the metric $g_{\mu\nu}$.  If we expand the metric tensor around some classical solution $g^{cl}_{\mu\nu}$ as
\be
g_{\mu\nu}=g_{\mu\nu}^{cl}+\sqrt{16\pi G}h_{\mu\nu}, 
\ee
so we can heuristically expand the Einstein-Hilbert action as 
\begin{align}\nonumber
\frac{1}{16\pi G}\int d^4\sqrt{-g}R=&\frac{1}{16\pi G}\int d^4 x\sqrt{-g^{cl}} R^{cl}\\\nonumber
&+\int d^4 x\sqrt{-g^{cl}}\Big(\partial h \partial h+Rh^2\\
&\qquad+\sqrt{16\pi G}h \partial h \partial h+16\pi G h^2\partial h \partial h+\ldots\Big),\label{gravexp}
\end{align}
where the terms in the second and third lines are really sums of similar terms with the indices contracted in various ways.  The first line of the right-hand side is independent of the metric perturbation $h_{\mu\nu}$ and thus does not affect the dynamics, the second line is the graviton kinetic term, and the third line describes graviton interactions suppressed by positive powers of the dimensionful coupling constant
\be
\ell_p\equiv \sqrt{\frac{8\pi G\hbar}{c^3}}\approx 8\times 10^{-35}\mathrm{m}.
\ee 
Here we display the factors of $c$ and $\hbar$, but from now on we will almost always set both to one.  In quantum field theory interactions which are suppressed by positive powers of such a length scale are \textit{non-renormalizable}, which means that the theory needs to be understood as a perturbative expansion in this length divided by the length scale of interest for the problem at hand, and moreover the higher-order terms in this expansion depend in detail on the short-distance properties of the theory.  For astrophysical black holes this ratio is quite small, of order $10^{-38}$ for stellar mass black holes and $10^{-47}$ for supermassive black holes, and so it is natural to first study only the leading order terms in this expansion where we replace $g_{\mu\nu}$ by $g_{\mu\nu}^{cl}$ and only treat $\phi$ quantum mechanically: this is called the approximation of quantum field theory in a fixed background metric.\footnote{Strictly speaking we should also include a free graviton controlled by the second line of \eqref{gravexp}, but this works in the same way as $\phi$ and they do not talk to each other so this is usually suppressed.}

A key feature of any quantum field theory is that there are fluctuations of the fields in the ground state, and moreover the fluctuations of fields which are near each other in space are strongly correlated.  We can see this already from the Hamiltonian of our free scalar in flat space:
\be\label{QFTH}
H=\frac{1}{2}\int d^3 x\left(\dot{\phi}^2+\vn\phi\cdot \vn \phi\right).
\ee
Here $\phi$ and $\dot{\phi}$ are canonical conjugates, and thus do not commute either with each other or with the Hamiltonian.  Since the ground state is an eigenstate of the Hamiltonian, there must be fluctuations of both $\phi$ and $\dot{\phi}$.  Moreover the fluctuations of $\phi$ at nearby points must be strongly correlated, since otherwise the $\vn\phi\cdot \vn \phi$ term in the Hamiltonian would lead to a large contribution to the energy.  The ground state of a quantum field theory is a pure quantum state, and so this correlation arises from entanglement.  In a nutshell, \textit{low-energy states of any quantum field theory are highly entangled}.  

\bfig
\includegraphics[height=5cm]{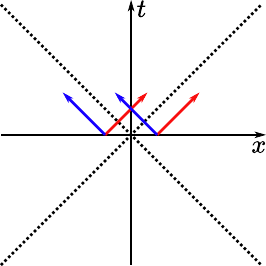}
\caption{Vacuum entanglement in a relativistic quantum field theory.  The blue left-moving modes on either side of $x=0$ are entangled with each other, as are the red right-moving modes.  This entanglement is necessary to avoid a large energy density, as explained below equation \eqref{QFTH}.}\label{rindlerfig}
\efig
In relativistic quantum field theories we can be more quantitative about this entanglement: every relativistic quantum field theory in flat space has a Lorentz boost operator 
\be
K_x=\int d^3x\left(xT_{00}+t T_{0x}\right), 
\ee
where $T_{\mu\nu}$ is the energy-momentum tensor, which is the generator of boosts 
\be
(t\pm x)'=(t\pm x)e^{\pm\lambda}
\ee
in the $tx$ plane (here $\lambda$ is called the ``rapidity'' of the boost).  Moreover we can define ``left'' and ``right'' boost operators via
\begin{align}\nonumber
K_x^R&=\int_0^\infty dx\int dy dz \left(xT_{00}+t T_{0x}\right)\\
K_x^L&= -\int_{-\infty}^0 dx\int dy dz \left(xT_{00}+t T_{0x}\right).
\end{align}
A standard path integral argument (see e.g. \cite{Harlow:2014yka}) then shows that ground state of the Hamiltonian $H$ is given by
\be
|\Omega\ran\propto \sum_i e^{-\pi \omega_i}|i^*\ran_L |i\ran_R,
\ee
where $|i\ran_R$ and $|i^*\ran_L$ are the eigenstates of $K_x^R$ and $K_x^L$ respectively, each with eigenvalue $\omega_i$. $|i\ran_R$ and $|i^*\ran_L$ are states of the fields in the regions $x>0$ and $x<0$, and ``$*$'' indicates that they are related by the action of an antiunitary operator called CRT which exists and commutes with the Hamiltonian in all relativistic field theories.\footnote{CRT implements the spatial reflection $x\to-x$, exchanges particles and antiparticles, and reverses time.  It is often combined with a rotation by $\pi$ in the $yz$ plane to produce another symmetry called $CPT$, and it is a theorem that both of these are a symmetry of any relativistic quantum field theory \cite{streater2016pct}.} This entanglement pattern is represented graphically in figure \ref{rindlerfig}.  We can also look at the reduced state on just the fields with $x>0$, which is given by the density operator
\be
\rho_R\propto e^{-2\pi K_x^R}.
\ee
Thus the fields of any quantum field theory in the region $x>0$ look thermal with respect to the right boost operator $K_x^R$, with all fluctuations arising from entanglement with the fields in the region $x<0$. The dimensionless ``temperature'' $\frac{1}{2\pi}$ may look somewhat puzzling, it is dimensionless because $K_x^R$ is.  In practice what it means is that in any relativistic quantum field theory, any observer undergoing a constant acceleration $a$ sees thermal fluctuations at a temperature \cite{Unruh:1976db,Unruh:1983ms}
\be
T_{Unruh}=\frac{a}{2\pi}=\frac{\hbar a}{2\pi k_B c},
\ee  
where in the second equality we have temporarily restored the dimensionful constants. 

\bfig
\includegraphics[height=6cm]{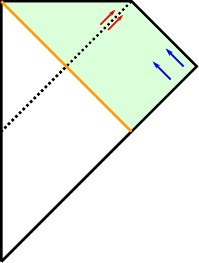}
\caption{Quantum field theory modes near a black hole formed from collapse.  The red right-moving modes on either side of the horizon must be entangled to avoid a large energy density there, but the blue left-moving modes should be in the ground state since nothing is being thrown into the black hole.}\label{hawkingmodesfig}
\efig
We can now consider what this vacuum entanglement says about black holes \cite{Hawking:1975vcx}. The time-translation symmetry $t'=t+b$ acts on the Kruskal coordinates \eqref{Kruskaldef} as
\be
X^{\pm'}=X^\pm e^{\pm \frac{b}{2r_s}},
\ee
so in other words it acts as a boost by the dimensionless rapidity $\frac{b}{2r_s}$. Therefore the relationship between the Hamiltonian $H$ generating $t$ translation and the boost generator $K$ near the Schwarzschild horizons is 
\be
H=\frac{K}{2 r_s}.
\ee
We just saw that in order to avoid a large energy density we need the fields on either side of the horizon to be entangled in such a way that the fields on either side look thermal with a temperature $\frac{1}{2\pi}$ relative to $K$.  In the Schwarzschild geometry we therefore need the fields to be entangled such that the temperature with respect to the Schwarzschild Hamiltonian $H$ is
\be\label{Thawk}
T_{Hawking}=\frac{1}{4\pi r_s}=\frac{1}{8\pi G M}=\frac{\hbar c^3}{8\pi k_B GM},
\ee
where we have again restored the dimensionful constants in the second equality.  In the Schwarzschild geometry this entanglement involves both left-moving and right-moving modes, so there is no net flux of energy.  The great insight of Hawking however was that this changes once we consider a genuine black hole created from collapse, such as that shown in figure \ref{penrosecollapsefig}.  Let's consider field modes on either side of this horizon at late times (well past the collapsing shell).  The basic situation is shown in figure \ref{hawkingmodesfig}: the right-moving modes just outside of the horizon need to be entangled with the right-moving modes just inside the horizon, just as in the Schwarzschild geometry, but the left-moving modes outside of the horizon no longer have this entanglement since their would-be partners have been removed by the presence of the infalling shell.  Indeed the left-moving modes just describe whatever objects we may have thrown into the black hole, and the simplest situation is therefore when we assume they are in their ground state (this is called the ``Unruh vacuum'').  Therefore near a black hole which has been left to itself, only the right-moving modes outside of the black hole are in a thermal state, and they are in precisely the thermal state with temperature $T_{Hawking}$.  

The statement that the right-moving (or ``outgoing'') quantum fields just outside of a black hole are thermal with temperature $T_{Hawking}$ while the left-moving (or ``ingoing'') quantum fields just outside are in their ground state has a remarkable consequence: it tells us that black holes should really be thought of as thermal objects possessing large numbers of microstates \cite{Bekenstein:1973ur,Hawking:1975vcx}.  In fact we can compute their entropy: using \eqref{Thawk} and
\be
\frac{dS}{dE}=\frac{1}{T},
\ee
and also assuming that the entropy of a zero mass black hole is zero, we have
\be\label{SBH}
S_{BH}=\frac{4\pi r_s^2}{4G}=\frac{\mathrm{Area}}{4G}.  
\ee
Here ``BH'' can either stand for black hole or Bekenstein-Hawking.  \eqref{SBH} is an extraordinary formula, in particular it says that the number of degrees of freedom in a spacetime region is proportional to the surface area of this region rather than its volume.  Note however that although were able to come up with a formula for $S_{BH}$, we have so far said nothing about what the fundamental degrees of freedom that realize this large number of microstates.  We will return to both of these points soon.  

Another remarkable consequence of the thermal state of outgoing modes near black holes is that it causes black holes to radiate and gradually lose mass.  The calculation of the evaporation rate is complicated by the fact that a black hole isn't a perfect blackbody, since incident radiation has some chance of being reflected before reaching the horizon, but up to order-one factors the evaporation rate is still given by the Stefan-Boltzmann formula
\be
\frac{dM}{dt}\sim-(4\pi r_s^2)T_{Hawking}^4\sim  -\frac{1}{G^2M^2}.
\ee
The time dependence of the mass is therefore given by
\be
M(t)=\left(M_0^3-\frac{Ct}{G^2}\right)^{1/3},
\ee
where $M_0$ is the initial mass and $C$ is an $O(1)$ number which depends on details like the number of light matter fields and what their spins are.  In particular the time it takes the black hole to fully evaporate is of order
\be\label{tevap}
t_{evap}\sim G^2 M_0^3,
\ee
which as mentioned in the introduction is a very long time for astrophysical black holes.

The results \eqref{SBH} and \eqref{tevap} for the entropy and lifetime of a black hole lead to a profound paradox \cite{Hawking:1975vcx,Hawking:1976ra}.  Say that we create a black hole in a pure quantum state $|\psi\ran$, perhaps using a spherical shell of photons as in figure \ref{hawkingmodesfig}.  As the mass gradually decreases, more and more energy is carried away by the Hawking radiation.  But according to the picture we have just discussed, this radiation is produced in a mixed state since it arises from the entanglement between the red modes in figure \ref{hawkingmodesfig}.  One way to quantify this is to note that the von Neumann entropy of the radiation is increasing as a function of time, as more and more thermal quanta escape from the black hole.  As we discussed around equation \eqref{gravexp}, this description is supposed to be accurate as long as $M(t)\gg \ell_p^{-1}$.  Therefore if we begin with a large black hole then almost all the energy in the radiation is produced during the period where the radiation entropy is increasing.  Once we reach $M(t)\sim \ell_p^{-1}$ the calculation is no longer reliable, but assuming energy conservation there are only two options for what happens next:
\begin{itemize}
\item[(1)] \textbf{Remnants:} The evaporation stops at $M\sim \ell_p^{-1}$, leaving behind a stable object, a ``remnant'', with a mass of order $\ell_p^{-1}$ and a very large number of microstates.  In fact the number of microstates needs to be infinite, since we could have started with an arbitrarily massive black hole. 
\item[(2)] \textbf{Information loss:} The evaporation completes, with the black hole disappearing in a final burst of radiation.  This final burst is not nearly entropic enough to purify the radiation which came out earlier, so the final state of the evaporation is mixed.  
\end{itemize}
Both of these options are quite unpleasant, since either one leads to either a violation of quantum mechanics or a situation where the entropy of black holes is not actually given by the Bekenstein-Hawking formula \eqref{SBH}.  Indeed if quantum mechanics is correct, then the combined state of the radiation produced so far and the black hole must be pure since the initial state was.  And in any pure state of a bipartite quantum system, the von Neumann entropy of one part is always equal to the von Neumann entropy of the other part.  Therefore the entropy of the remaining black hole must always equal that of the Hawking radiation.  But since the entropy of the Hawking radiation is increasing as a function of time, while the size of the black hole is decreasing, eventually the former will exceed the area of the horizon in Planck units of the latter.  This is only possible of the number of microstates of the latter is not actually given by the exponential of $A/(4G)$.  Why then does the black hole radiate at a temperature $T_{Hawking}$?  

Given the unsatisfactory nature of options (1-2), most black hole theorists today instead believe in a third option:
\begin{itemize}
\item [(3)] \textbf{Unitary evaporation:} Hawking's calculation of black hole radiation is only correct in a coarse-grained sense: the detailed state of the hawking radiation cannot be computed even when $M(t)\gg \ell_p^{-1}$, and in a true theory of quantum gravity the final state of the radiation is actually pure.  Moreover this final state is given by a unitary transformation of the initial state, just as in ordinary quantum mechanics.  
\end{itemize}
This option may seem to obviously be preferable to the previous two, but in fact it is also quite radical: it requires a large violation of locality in a situation where the effective field theory \eqref{gravexp} seems like it should be accurate to an excellent approximation.  The scale of non-locality which is required is apparent in figure \ref{hawkingmodesfig}: at late times the information about the initial shell is deep inside the black hole but some kind of interaction needs to move it out into the red Hawking radiation even though the two are spacelike separated.  

\bfig
\includegraphics[height=4cm]{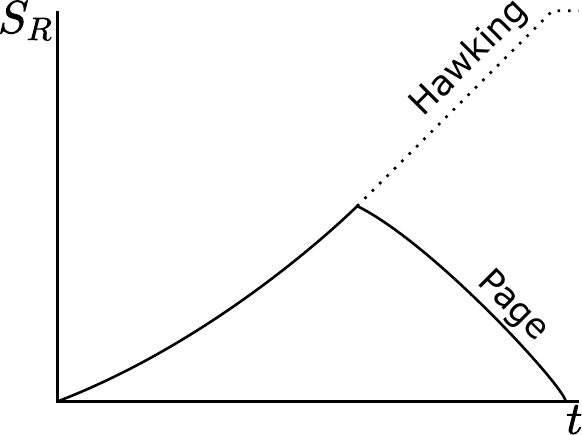}
\caption{Possible plots of the radiation entropy $S_R$ as a function of time, usually called the Page curve, for an evaporating black hole: Hawking's calculation based on effective field theory suggests that the final entropy of the radiation is mixed, while unitarity says it should be pure.}\label{pageoptionsfig}
\efig
There is a useful way to illustrate the distinctions between options (1-3): we study the von Neumann entropy of the Hawking radiation as a function of time, which these days is called the \textit{Page curve} \cite{Page:1993wv}.  The possibilities are shown in figure \ref{pageoptionsfig}.  Note in particular that any modification which leads to unitary evaporation must lead to a large modification of $\frac{dS_R}{dt}$ away from Hawking's prediction, and it must achieve this in the regime while $M\gg \ell_p^{-1}$: this is another way of seeing that unitarity requires a substantial breakdown of effective field theory \cite{Mathur:2009hf}.  

Thus we see that Hawking's black hole information paradox requires us to make a choice which is both terrible and wonderful: we need to decide whether to give up quantum mechanics, statistical mechanics, or locality.  All of these are core principles of modern physics, and so we are guaranteed to learn something novel by resolving the paradox.  The mainstream view these days is that locality is most likely the culprit, and the rest of this article will give some overview of why.  Here is a list of what I view as the main advantages of option (3):
\begin{itemize}
\item In a theory with unitary black hole evaporation black holes behave just like any other complex quantum systems, and in particular they obey laws of thermodynamics which are statistical in origin just as gases or materials do.  
\item Already using semiclassical physics we can estimate the potential amount of information which can be stored in the Hawking radiation of a black hole, and there is indeed enough ``room'' to keep track of the initial state.  Very roughly, we can think of the Hawking cloud produced by a black hole of initial mass $M$ as a $1+1$ dimensional thermal gas of temperature $1/GM$ and size $G^2 M^3$.\footnote{It is $1+1$ dimensional because the radiation is dominated by low-angular momentum modes, since those with higher angular momentum are very unlikely to make it out of the black hole.}  Thus coarse-grained thermal entropy is
\be
S_{cg}\sim \frac{1}{GM}\times G^2 M^3\sim GM^2\sim \frac{(GM)^2}{G}\sim \frac{\mathrm{Area}}{G},
\ee
so unitarity is compatible with the coarse-grained properties of the system.  
\item In string theory, which is our best theory of quantum gravity so far, we are directly able to count the microstates of some special black holes and we find that the number is indeed the exponential of $\frac{\mathrm{Area}}{4G}$.  As just discussed, this is incompatible with remnants or information loss.
\item Within the ``AdS/CFT correspondence'', a corner of string theory over which we have especially good control, we have recently learned how to reliably compute the Page curve for some special evaporating black holes.  The answer is consistent with Page's proposal in figure \ref{pageoptionsfig}, and moreover the method seems likely to generalize to any evaporating black hole.  
\end{itemize}
The remainder of this article will discuss these last two points in more detail.

\section{Black holes in string theory}
\bfig
\includegraphics[height=4cm]{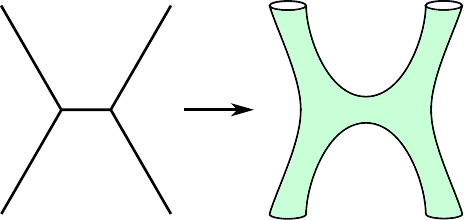}
\caption{In string theory Feynman diagrams are resolved into smooth surfaces, resulting in amplitudes which are well-defined order by order in $E\sqrt{G}$.  Here $E$ is the energy of the process being considered.}\label{stringscattfig}
\efig
String theory is a mathematical formalism which describes the interactions of relativistic strings.  It was originally proposed as a theory of hadrons, whose structure at low energies is characterized by string-like interactions between quarks and gluons.  These strings were soon understood to be tubes of nonabelian gauge flux however, and it turned out to be more useful to organize the theory of the strong force in terms of this flux, leading to the development of QCD.  It was eventually realized  that string theory works better as a theory of quantum gravity \cite{Scherk:1974ca,Green:1984sg}.  The natural relativistic string action is  
\be
S=-\frac{1}{2\pi \ell_s^2}\int d^2 x \sqrt{-\gamma}\left(1+\frac{\Phi_0\ell_s^2}{2}R\right),
\ee
where $\sqrt{-\gamma}$ is the induced volume element on the two-dimensional spacetime surface swept out by the string, usually called the worldsheet, and $R$ is the Ricci scalar for this induced metric.  $\frac{1}{2\pi \ell_s^2}$ is the string tension, and $\ell_s$ is called the string length and sets the typical size of the strings.  For the QCD string we have $\ell_s\sim 10^{-15} \mathrm{m}$, while for quantum gravity $\ell_s$ is much smaller and is often taken to be of order $10^{-33}\mathrm{m}$.  The meaning of the parameter $\Phi_0$ is that the quantity 
\be\label{gdef}
g\equiv e^{\Phi_0}
\ee
plays the role of a coupling constant: it suppresses higher topologies of the worldsheet, and these higher topologies are precisely what mediate interactions between strings.  Figure \ref{stringscattfig} gives the simplest example of a string interaction. Loop diagrams are surfaces with more handles, and thus are suppressed by higher powers of $g$. 

One very nice feature of string theory as a theory of quantum gravity is that gives a way to resolve the non-renormalizibility of Einstein gravity.  This happens because the interactions are ``spread out'' in spacetime instead of localized at points as they would be in the effective theory \eqref{gravexp}.  The theory thus has a well-defined perturbation theory to all orders in Newton's constant.  On the other hand it comes with two substantial disadvantages: in order to preserve Lorentz invariance and the stability of the vacuum it is necessary for these strings to propagate in ten spacetime dimensions instead of four, and it is also necessary that they carry additional fermionic degrees of freedom which make the theory supersymmetric (supersymmetry is a symmetry exchanging bosons and fermions).  These results at first seem to immediately falsify string theory as a theory of our world, since we don't live in ten dimensions and nature does not look supersymmetric, but the truth is more subtle: string theory has no continuous parameters, and is thus is in some sense unique, but its low-energy description does have fields whose expectation values can be varied continuously by changing the initial state of the universe ($\Phi_0$ is such an expectation value), and these expectation values can be used to break supersymmetry and curl up some of the ten spacetime dimensions to very small size - a process called \textit{compactification}.  Since there are many different ways to curl up the extra six dimensions and break supersymmetry, string theory has a large ``landscape'' of solutions with four large spacetime dimensions, and each of these solutions, sometimes called vacua, gives rise to different particles and interactions at low energies.  

Looking for string vacua with realistic particle physics is a difficult technical problem, which so far has not been completely solved, but to study black hole physics in string theory this problem does not need to be solved.  Indeed the black hole information problem is just as severe in supersymmetric theories and in higher dimensions, so we are free to pick whichever compactification is most convenient.  Before doing this however, we need to discuss the ingredients of string theory in a bit more detail.  We'll begin directly in ten dimensions, where no compactification has yet happened.  It turns out that at first there seem to be five distinct possible choices for how to construct a ten-dimensional string theory, which are called IIA, IIB, IA, $SO(32)$ heterotic, and $E_8\times E_8$ heterotic.  For brevity we will mostly just discuss IIA and IIB. We are particularly interested in their low-energy limits, which are described by IIA and IIB supergravity respectively.  IIA supergravity is a ten-dimensional gravity theory whose fields are a metric $g_{\mu\nu}$, a scalar $\Phi$ called the dilaton, a two-form gauge field $B_{\mu\nu}$ called the Kalb-Ramond field, two ``Ramond-Ramond'' $p$-form gauge fields $A_{\mu}$, and $\wt{A}_{\mu\nu\lambda}$, and various fermionic partners of these bosonic fields.  The action is
\begin{align}\nonumber
S=&\frac{1}{2\kappa_{10}^2}\int d^{10}x \sqrt{-g}\Bigg[e^{-2\Phi}\left(R+4\partial_\mu \Phi\partial^\mu \Phi-\frac{1}{24}H_{\mu\nu\sigma}H^{\mu\nu\sigma}\right)\\\nonumber
&\qquad -\frac{1}{4}F_{\mu\nu}F^{\mu\nu}-\frac{1}{48}\wt{F}_{\mu\nu\rho\sigma}\wt{F}^{\mu\nu\rho\sigma}\Bigg]\\\nonumber
&-\frac{1}{4\kappa_{10}^2}\int B\wedge \wt{F}\wedge \wt{F}\\\label{IIAS}
&+\ldots,
\end{align}
where ``$\ldots$'' indicates terms involving the fermionic fields and $H=dB$, $F=dA$, and $\wt{F}=d\wt{A}$ where $d$ is the exterior derivative.  The Newton-like constant $\kappa_{10}^2$ is related to the string length $\ell_s$ by
\be
2\kappa_{10}^2=(2\pi)^7\ell_s^8,
\ee
and the expectation value of the dilaton is precisely the quantity we were previously calling $\Phi_0$.  The easiest way to remember all these fields is to realize that the IIA theory can be obtained as the dimensional reduction of an \textit{eleven-dimensional} supergravity theory on a small spatial circle: this eleven-dimensional theory has only two bosonic fields, a metric $g_{\mu\nu}$ and a three-form gauge field $M_{\mu\nu\sigma}$.  The Lagrangian is given by
\be\label{Mact}
S=\frac{1}{2\kappa_{11}^2}\int d^{11}x \sqrt{-g}\left(R-\frac{1}{48} T_{\alpha\beta\rho\sigma}T^{\alpha\beta\rho\sigma}\right)-\frac{1}{12\kappa_{11}^2}\int M\wedge T\wedge T+\ldots,
\ee
where $T=dM$ and we have again neglected terms involving fermions.  The IIA supergravity fields then arise as follows: $e^{2\Phi/3}$ is the radius of the compactified $S^1$, $A$ is the ``Kaluza-Klein'' gauge field arising from off-diagonal components of the metric, $B$ arises from $M$ with one index on the $S^1$, and $\wt{A}$ arises from $M$ with all indices in the noncompact directions.  IIB supergravity is governed by a similar action, except instead of the Ramond-Ramond fields $A$ and $\wt{A}$ we instead have three Ramond-Ramond $p$-form fields $C$ (a zero-form), $C_{\mu\nu}$, and $C_{\mu\nu\rho\sigma}$. 

This connection between 11 dimensional supergravity and IIA supergravity is not a coincidence: the strongly-coupled limit $\Phi_0\to \infty$ of IIA string theory seems to be given by some mysterious eleven-dimensional theory called ``M-theory'' whose low-energy limit is indeed controlled by the action \eqref{Mact}.  And moreover IIA and IIB string theory are related in the following way: if we compactify the IIA theory on a circle of radius $r_A$, then in the limit where $r_A\ll \ell_s$ then the theory is actually to the IIB theory compactified on a circle whose radius $r_B$ is \textit{large} compared to the length scale of $\kappa_{10}$. More concretely we have
\be
r_B=\frac{\ell_s^2}{r_A},
\ee  
which is a relation known as T-duality.  The other three string theories are also related to the IIA and IIB theories along similar lines, and so there seems to really only be a single non-perturbative theory which is described by the various perturbative string theories in different limits \cite{Witten:1995ex}.

Given the various gauge fields in the IIA and IIB theories, it is natural to ask if there are objects which are charged under these gauge fields.\footnote{In fact in quantum gravity it is widely expected that charged objects \textit{must} exist for any gauge field.  This topic is beyond the scope of this article, see \cite{Polchinski:2003bq,Banks:2010zn,Harlow:2018tng} for more on this.}  In general the objects which are charged under a $p$-form gauge field are objects with $p-1$ spatial dimensions, so for example an ordinary one-form gauge field $A_\mu$ can be integrated along the worldline of a charged particle to give a natural action
\be
S=q_1\int A.
\ee
Here $q_1$ is the charge of the particle.  More generally we can integrate a $p$-form gauge field $A$ over the worldvolume of an object with $p-1$ spatial dimensions.  In IIA string theory the strings themselves are the objects which are charged under the Kalb-Ramond two-form $B_{\mu\nu}$, but it is less clear what objects should be charged under $A_{\mu}$ and $\wt{A}_{\mu\nu\sigma}$.  Such charges do in fact exist, and they are called D0 and D2 branes respectively.  In string perturbation theory they are objects on which IIA strings can end.  Similarly in the IIB theory there are D1, D3, and D5 branes which are charged under $C$, $C_{\mu\nu}$, and $C_{\mu\nu\rho\sigma}$ respectively \cite{Polchinski:1995mt}.  D-branes are sources for the metric as well as the RR gauge fields, and the classical solution of the IIA/IIB supergravity equations which is sourced by $N$ $Dp$ branes is 
\begin{align}\nonumber
ds^2&=\frac{1}{\sqrt{Z_p(r)}}\eta_{\mu\nu}dx^\mu dx^\nu+\sqrt{Z_p(r)}\left(dr^2+r^2d\Omega^2_{8-p}\right)\\\nonumber
e^{2\Phi}&=g^2Z_p(r)^{\frac{3-p}{2}}\\
A_{p+1}&=g^{-1}\left(\frac{1}{Z_p(r)}-1\right)dx^0\wedge \ldots \wedge dx^p.\label{pbranes}
\end{align}
Here $A_{p+1}$ is whichever RR gauge field is sourced by the brane, $g$ is the string coupling \eqref{gdef}, and 
\be
Z_p(r)\equiv 1+\frac{(4\pi)^{\frac{5-p}{2}}\Gamma\left(\frac{7-p}{2}\right)gN\ell_s^{7-p}}{r^{7-p}}.
\ee

\bfig
\includegraphics[height=5cm]{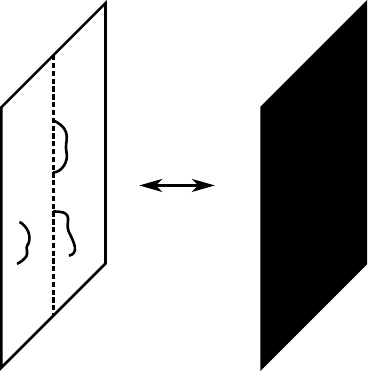}
\caption{A D-brane black hole in string theory.  At weak coupling we have a stack of D1 branes (the dashed line) living inside of a stack of D5 branes, with excitations described be weakly-interacting strings whose endpoints can be attached to either kind of brane.  At strong coupling this system turns into a black brane, which looks like a black hole to a low-energy observer since the spatial extent of the brane is all in the compact directions.}\label{stringbhfig}
\efig
What can we learn about black holes from string theory?  So far there are two main achievements.  The first is that within string theory we are able to directly count the microstates of certain supersymmetric black holes, and the results agree in all cases with the Bekenstein-Hawking formula \eqref{SBH} \cite{Strominger:1996sh}.  There are many examples, but one particularly simple one consists of a black hole which is created by compactifying IIB string theory on a five-dimensional torus, wrapping $N_1$ D1 branes and $N_5$ D5 branes on this torus, giving the system a momentum 
\be
p_5=\frac{2\pi m_5}{L}
\ee 
in the direction of the torus cycle that the D1 branes are wrapping (which we take to have length $L$), and working at zero temperature.  If we fix the value of the string coupling constant $g$ and take $N_1$, $N_5$, and $m_5$ to all be large then this system becomes a five-dimensional supersymmetric black hole, whose geometry and matter fields are a more complicated version of \eqref{pbranes}.  The Bekenstein-Hawking entropy of this black hole is\cite{Polchinski:1998rr}
\be\label{stringbh}
S=\frac{\mathrm{Area}}{4G}=\frac{2\pi \mathrm{Area}}{g^2\kappa_{10}^2}=2\pi \sqrt{N_1 N_5 m_5}.
\ee
On the other hand we can take $N_1$, $N_5$, and $m_5$ to be large but fixed and then send $g\to 0$.  In this limit the system is described by a bunch of free strings attached to the D-branes (see figure \ref{stringbhfig}).  We are interested in counting the number of states with momentum $p_5$ in the D1-brane direction, which at low temperatures controlled by modes which depend only on time and the D1-brane direction.  The dynamics of these modes is described by a two-dimensional conformal field theory, whose density of states at large $p_5$ can be computed using standard techniques.  The result is that at large charge the density of states is\footnote{A sketch of the counting goes like this: there are $4N_1^2$ scalar degrees of freedom arising from strings attaching the D1 to itself, $4 N_5^2$ scalar degrees of freedom arising from strings attaching the D5 to itself, and $4N_1 N_5$ degrees of freedom arising from strings attaching the D1 to the D5.  The potential for these scalars makes $3(N_1^2+N_2^2)$ of them massive, and we subtract a further $N_1^2+N_5^2$ due to quotienting by the $U(N_1)\times U(N_5)$ gauge symmetry on the branes.  There are thus $4 N_1 N_5$ massless scalar degrees of freedom. By supersymmetry these must have $4N_1 N_5$ fermionic partners, leading to a left-moving central charge $c_L=6 N_1 N_5$ for the 2D CFT (fermions only contribute $1/2$ to the central charge).  We are interested in counting states with left-moving energy $E=p_5=\frac{2\pi m_5}{L}$ (states with both left and right moving energy are suppressed at low temperature), and we can use the ``Cardy formula'', which tells us that at high energy the density of left-moving states of energy $E$ for any 2D CFT on a spatial circle of length $L$ is $\rho \approx e^{\sqrt{\frac{\pi c_L LE}{3}}}$.}
\be
\rho\approx e^{2\pi \sqrt{N_1 N_5 m_5}},
\ee
in perfect agreement with \eqref{stringbh}.  At first this agreement may seem accidental, since we counted the black hole entropy at fixed $g$ and large charge while we counted the D-brane entropy at fixed charge and small $g$.  This however is where supersymmetry comes to the rescue: it turns out that these states are all ``BPS'', which means that they have the minimal energy for a given charge which is allowed by the supersymmetry algebra.  Such states transform in smaller representations of the supersymmetry algebra than do states which do not saturate this bound, and therefore the number of BPS states cannot change as we vary a continuous parameter $g$ unless there are other BPS states around with the same charges which they can pair up with.  No such states exist here, so we can reliably extend the D-brane entropy counting at weak coupling into the black hole regime, where the agreement with \eqref{stringbh} can indeed be viewed as a test of the Bekenstein-Hawking formula.  

The second achievement of string theory related to black hole physics is the discovery of the AdS/CFT correspondence, which says that quantum gravity in asymptotically-AdS space is non-perturbatively equivalent to conformal field theory living on the boundary of that space.  This correspondence was discovered in string theory in the following way.  Consider a stack of $N$ D3 branes.  The solution \eqref{pbranes} becomes 
\begin{align}\nonumber
ds^2&=\frac{1}{\sqrt{1+\left(\frac{\ell_{ads}}{r}\right)^4}}\eta_{\mu\nu}dx^\mu dx^\nu+\sqrt{1+\left(\frac{\ell_{ads}}{r}\right)^4}\left(dr^2+r^2 d\Omega_5^2\right)\\\nonumber
e^{2\Phi}&=g^2\\
C_4&=-g^{-1}\frac{\ell_{ads}^4}{\ell_{ads}^4+r^4}dx_0\wedge\ldots \wedge dx^4,\label{D3}
\end{align}
with 
\be\label{adsscale}
\ell_{ads}^4 \equiv 4\pi g N \ell_s^4.
\ee
The curvature radius of this geometry is of order $\ell_{ads}$, so the dynamics of strings near the stack qualitatively depend on whether $gN$ is large or small.  If $gN\ll 1$ then $\ell_{ads}\ll \ell_s$ so the curvature all vanishes below the string scale and we can think of the branes as background objects in ten-dimensional flat space.  Their excitations are described by weakly interacting strings, just as in the left picture in figure \ref{stringbhfig}.  When $gN\gg 1$ then this picture of weakly-interacting strings attached to the branes is not valid, but there is an alterative description which is: we now have $\ell_{ads}\gg \ell_s$, so the geometry \eqref{D3} gives a good semiclassical description of what is going on even in the region near the branes where $r\ll \ell_{ads}$ and the states consist of closed strings moving in this geometry.  Moreover the geometry simplifies in the region $r\ll \ell_{ads}$: we have
\be
ds^2\approx \left(\frac{r}{\ell_{ads}}\right)^2\eta_{\mu\nu} dx^\mu dx^\nu+\left(\frac{\ell_{ads}}{r}\right)^2dr^2+\ell_{ads}^2d\Omega_5^2,\label{ads5}
\ee
which is precisely the geometry of $AdS_5\times \mathbb{S}^5$, where $AdS_5$ is the maximally symmetric solution of Einstein's equations in five spacetime dimensions with negative cosmological constant.  The parameter $\ell_{ads}$ sets the radius of this negative curvature, and also the size of the $\mathbb{S}^5$.  The great insight of Maldacena \cite{Maldacena:1997re} is that at low energies we actually know the dynamics of the D3-brane stack at \textit{any} value of $gN$: it is always given by the maximally-supersymmetric Yang-Mills theory in $3+1$ dimensions with gauge group $U(N)$ and Yang-Mills coupling 
\be\label{grel}
g^2_{YM}=4\pi g.
\ee
This theory has a $U(N)$ gauge field $A_\mu$, six real scalars $X_i$, and four Majorana fermions $\psi_a$, all transforming in the adjoint of $U(N)$.  The Lagrangian density is
\be\label{N4}
\mathcal{L}=-\frac{1}{2g_{YM}^2}\Tr\left(F_{\mu\nu}F^{\mu\nu}\right)-\sum_i \Tr\left(D_\mu X_i D^\mu X_i\right)+\frac{g_{YM}^2}{2}\sum_{i,j}\Tr\left([X_i,X_j]^2\right)+\ldots,
\ee
where $A_\mu$ and $X_i$ are both $N\times N$ matrices and $\ldots$ indicates kinetic and Yukawa terms for the fermions $\psi_a$.  This statement can be checked at small $gN$ using string perturbation theory in the vicinity of the branes, and in fact the large amount of supersymmetry ensures that it then holds for all values of $gN$ since there is no other Lagrangian which can be written down with the necessary symmetries.  Since at large $gN$ we've seen that the dynamics near the branes are described by IIB string theory in asymptotically $AdS_5\times \mathbb{S}^5$ space, it is natural to propose that these two descriptions are equivalent even at finite $gN$: the non-perturbative formulation of IIB string theory in asymptotically $AdS_5\times \mathbb{S}^5$ spacetime is nothing but the maximally supersymmetric Yang Mills theory with action \eqref{N4}!  This proposal has several remarkable features:
\bi
\item The gauge theory description \eqref{N4} of the system is a non-gravitational quantum field theory, so we have reformulated something apparently mysterious, quantum gravity, in terms of something much more conventional.  In particular the quantum field theory description obeys all the usual axioms of quantum mechanics, in particular its time evolution is unitary.  
\item From the first line of \eqref{IIAS} (which is the same as in the IIB Lagrangian), the ten-dimensional Planck length is
\be
\ell_{10}^8\equiv g^2\kappa_{10}^2=\frac{g^2}{2}(2\pi)^7\ell_s^8.
\ee
Therefore the AdS radius \eqref{adsscale} in Planck units is of order
\be
\frac{\ell_{ads}}{\ell_{10}}\sim N^{1/4}.
\ee
From \eqref{adsscale} and \eqref{grel} we also have
\be
\frac{\ell_{ads}}{\ell_s}\sim (g_{YM}^2 N)^{1/4},
\ee
so in the semiclassical limit where the gravitational picture is well-described by low-energy effective field theory we have
\begin{align}\nonumber
N&\gg 1\\
g_{YM}^2N&\gg 1.
\end{align}
The limit $N\gg 1$ is very well-studied in gauge theories: in the 1970s 't Hooft predicted that they should become semiclassical in this limit \cite{tHooft:1973alw}, and in the maximally-supersymmetric case with $g_{YM}^2N\gg 1$ we now see what this semiclassical theory is - IIB supergravity in $AdS_5 \times \mathbb{S}^5$!\footnote{More generally if $N$ is large but $g_{YM}^2N$ is not then we have classical IIB string theory in $AdS_5 \times \mathbb{S}^5$.}
\item IIB supergravity has an interesting ``S-duality'' operation which sends $\Phi\to -\Phi$, $g_{\mu\nu}\to e^{-\Phi} g_{\mu\nu}$, $C_{\mu\nu}\to -B_{\mu\nu}$, and $B_{\mu\nu}\to -C_{\mu\nu}$, leaving the action invariant. This operation inverts the string coupling $g\to 1/g$, and thus from a string theory point of view is an operation which exchanges strong and weak coupling.  Moreover there are good indications that this operation preserves the full IIB string theory and not just supergravity.  If this is the case, then it must also be a feature of the maximally-supersymmetric gauge theory \eqref{N4}.  And this indeed seems quite likely to be true: there are many pieces of evidence that the theory \eqref{N4} with gauge group $U(N)$ is the same theory with Yang-Mills coupling $\frac{4\pi}{g_{YM}}$ as it is with Yang-Mills coupling $g_{YM}$.
\ei

\section{Holography, black holes, and the emergence of spacetime}
The relationship between IIB string theory in $AdS_5\times S^5$ and the supersymmetric gauge theory \eqref{N4} described in the last section is now understood as an example of a much more general phenomenon: the AdS/CFT correspondence.  In this final section we will study this correspondence as a logical framework in its own right, without further discussion of its situation within string theory (although we emphasize that string theory remains the only known way of constructing examples of the correspondence which have a sizable hierarchy between the Planck/string scales and the AdS scale).  In particular the AdS/CFT correspondence gives us our first example of a well-defined theory of quantum gravity which is realistic enough to have all the ingredients of the black hole information problem, so we will use it to study that paradox, seeing that we obtain results which are consistent with unitary evaporation.\footnote{In low numbers of spacetime dimensions there are renormalizable theories of gravity which either do not have black holes or do not have propagating degrees of freedom.  None of these is rich enough to include all the ingredients of Hawking's paradox.}  

\bfig
\includegraphics[height=5cm]{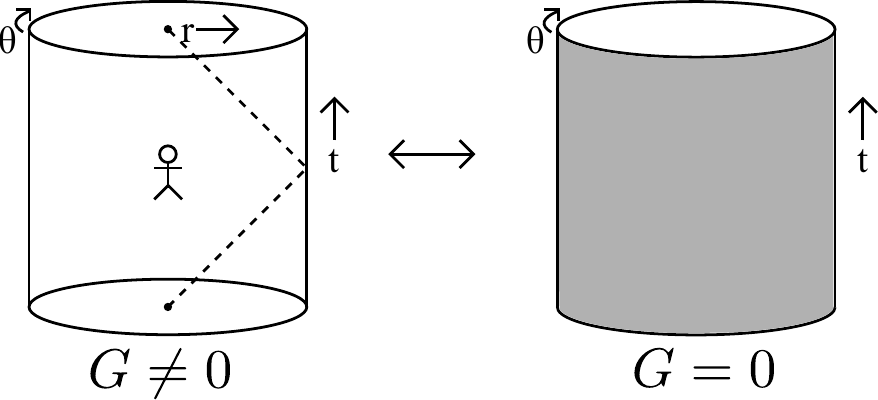}
\caption{The AdS/CFT correspondence.  On the left we have gravity in asymptotically-AdS space, which is represented as the interior of the cylinder.  The dashed line represents a photon which travels out to the boundary and returns in a finite amount of proper time as seen by the person sitting in the center.  The fundamental description of this situation is shown on the right: a non-gravitational quantum field theory living at the asymptotic boundary, which is shaded grey.}\label{adscftfig}
\efig
Formally, the AdS/CFT correspondence says that quantum gravity on the set of spacetimes which asymptotically approach $AdS_d$ is equivalent to conformal field theory on a spatial $\mathbb{S}^{d-1}$.  To explain this further, it is convenient to write the $AdS_d$ metric as (these are different coordinates than the ones used in \eqref{ads5})
\be\label{adsg}
ds^2=-\left(1+\frac{r^2}{\ell_{ads}^2}\right)dt^2+\frac{dr^2}{1+\frac{r^2}{\ell_{ads}^2}}+r^2 d\Omega_{d-2}^2.
\ee
This geometry resembles that of flat space for $r\ll \ell_{ads}$, while at distances which are large compared to $\ell_{ads}$ it tends to pull objects back towards $r=0$ (this can be thought of as the opposite of the accelerating expansion of the universe in dS space).  The asymptotic boundary where we can think of the dual CFT as living is at $r\to \infty$.  See figure \ref{adscftfig} for a basic illustration of the correspondence.  The word ``conformal'' here refers to the fact that the metric \eqref{adsg} has a large isometry group $SO(d-1,2)$, which arises from the fact that this metric can be obtained as the induced metric on a hyperboloid 
\be
T_1^2+T_2^2-\vec{X}^2=\ell_{ads}^2
\ee
embedded in a $(d+1)$-dimensional Minkowski space with two timelike directions.  As a statement about the boundary quantum field theory, this says that in additional to the usual Poincare symmetry
\be
x^{\mu\prime}=\Lambda^\mu_{\phantom{\mu}\nu}x^\nu+a^\mu
\ee
the theory must also be invariant under the scaling symmetry 
\be
x^{\mu\prime}=\lambda x^\mu
\ee 
and the ``special conformal transformations''
\be
x^{\mu\prime}=\frac{x^\mu+b^\mu x^\nu x_\nu}{1+2 b_\alpha x^\alpha+b_\beta b^\beta x_\gamma x^\gamma}:
\ee 
quantum field theories with this symmetry are called \textit{conformal field theories}, and the supersymmetric Yang-Mills theory \eqref{N4} is indeed conformal. In conformal field theory local operators are classified by how the transform under these conformal symmetries, and a particularly interesting set of local operators are \textit{primary operators}, which obey
\begin{align}
e^{iD\alpha}O(x)e^{-iD\alpha}&=e^{\alpha \Delta}O(e^\alpha x)\\
e^{iK_\mu a^\mu}O(0)e^{-i K_\mu a^\mu}&=O(0),
\end{align}
where $D$ and $K_\mu$ are the generators of dilations and special conformal transformations respectively. 

For the AdS/CFT correspondence to be useful, we need a ``dictionary'' which tells us how observables in the gravity description are represented in the CFT description, and vice versa.  In general conformal field theories the dual gravitational description will not have a large hierarchy between the AdS scale $\ell_{ads}$ and the Planck scale $\ell_p$, in which case we can't say much about the dictionary: basically just the generators of $SO(d-1,2)$ should match on the two sides.  We are particularly interested however in the situation where such a hierarchy does exist, as it is only in this situation that the gravity description has any resemblance to the gravity in our world.  We therefore introduce the notion of a conformal field theory with a \textit{semiclassical dual}: this is a CFT with a finite set of primary operators $O_i$ and a bulk effective action $S(\phi_i,\Lambda)$, with $\phi_i$ a set of bulk fields including a metric $g_{\mu\nu}$, such that we have
\be
\int \mathcal{D}\phi_i e^{iS(\phi_i,\Lambda)}O_{i_1}^{bulk}(t_1,\Omega_1)\ldots O_{i_n}^{bulk}(t_n,\Omega_n)= \lan O_{i_1}(t_1,\Omega_1)\ldots O_{i_n}(t_n,\Omega_n)\ran_{CFT}\label{scdual}
\ee
to all orders in $\frac{1}{\ell \Lambda}$ (here $n$ is restricted to be $O((\frac{1}{\ell \Lambda})^0)$).  The bulk observables here are defined by the ``extrapolate dictionary'':
\be\label{extdict}
O_{i}^{bulk}(t,\Omega)\equiv \lim_{r\to\infty}r^{\Delta_i}\phi_i(r,t,\Omega).
\ee
Any CFT obeying this definition will be dual to a gravitational theory with ``reasonable particle physics''.  In particular two natural pairings between bulk fields and boundary operators are
\be
g_{\mu\nu}\leftrightarrow T_{\mu\nu}
\ee
and 
\be
A_\mu \leftrightarrow J_\mu,
\ee
where $g_{\mu\nu}$ and $A_\mu$ are the bulk metric and a bulk gauge field and $T_{\mu\nu}$ and $J_\mu$ are the boundary energy momentum tensor and a conserved current.  The supersymmetric gauge theory \eqref{N4} can be fairly explicitly shown to have a semiclassical dual in this sense, with the effective action $S$ being given by IIB supergravity and the primary operators $O_i$ given by traces of certain polynomials in the Super-Yang Mills fields which are chosen to preserve a large amount of supersymmetry.  It is worth emphasizing that the bulk description of the physics in terms of $S(\phi_i,\Lambda)$ is only approximate: it is not a good description in all states, and in some states it is a very bad description.  Another way to say this is that the bulk description is \textit{emergent}.  

\bfig
\includegraphics[height=4cm]{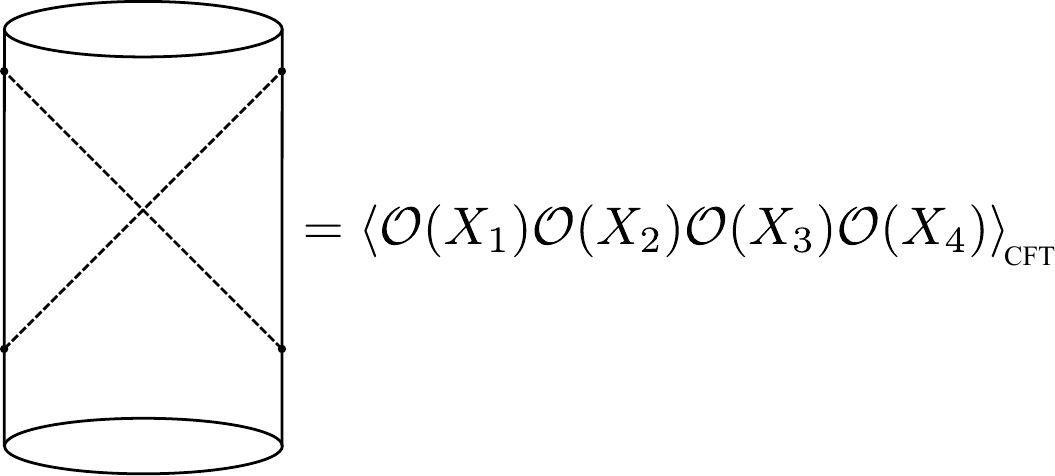}
\caption{Using boundary correlators to learn about bulk scattering in the AdS/CFT correspondence.}\label{ads4ptfig}
\efig
The easiest way to learn about gravitational physics from the dual CFT description is by way of the dictionary \eqref{extdict}: using correlation functions of local primary operators in the dual CFT we can conduct ``scattering experiments'' in bulk description, as shown in figure \ref{ads4ptfig}.  This includes simple ``particle physics'' scattering of small numbers of particles, but we can also study experiments such as Hawking's information loss experiment where we collapse some matter and then wait to see if the cloud of Hawking radiation which comes out is pure.  If the AdS/CFT correspondence is correct, then apparently it will be pure since the time evolution in the dual CFT is manifestly unitary.

\bfig
\includegraphics[height=5cm]{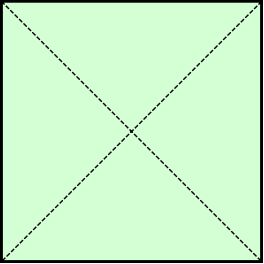}
\caption{Penrose diagram of the AdS-Schwarzschild geometry, with two asymptotically-AdS boundaries connected by a wormhole.}\label{penroseadsschfig}
\efig
In practice however this test of black hole unitarity is rather formal, in that we were rather implicit about how exactly the final state of the radiation should be determined from boundary correlation functions.  In fact there is a simpler test, based on looking at the long-time behavior of thermal two-point functions in the dual CFT \cite{Maldacena:2001kr}.  To understand this better, we first need to discuss in some more detail the AdS version of the Schwarzschild geometry, which has metric   
\be
ds^2=-f(r)dt^2+\frac{dr^2}{f(r)}+r^2 d\Omega_{d-2}^2
\ee
with
\be
f(r)=\frac{r^2}{\ell_{ads}^2}+1-\frac{16\pi G M}{(d-2)\Omega_{d-2}r^{d-3}}.
\ee
This metric approaches the AdS metric \eqref{adsg} at large $r$, while for $\ell_{ads}\to \infty$ and $d=4$ it becomes the usual Schwarzschild metric \eqref{schdef}.   For $M>0$ the function $f(r)$ has a unique positive zero $r_s$, at which there is a horizon just as for the usual Schwarzschild metric, and by introducing Kruskal coordinates analogous to \eqref{Kruskaldef} we can see that the full geometry again consists of two asymptotic universes connected by a wormhole.  The Penrose diagram is shown in the left diagram of figure \ref{penroseadsschfig}.  We can also construct single-boundary black hole solutions by collapsing a shell of matter just as in figure \ref{penrosecollapsefig}, and these black holes will produce Hawking radiation just as in flat space, but with a modified temperature
\be
T=\frac{(d-3)+(d-1)r_s^2/\ell_{ads}^2}{4\pi r_s}.
\ee  
In particular note that when $r_s\gg \ell_{ads}$, in which case the black hole is called ``big'', then the temperature increases with $r_s$, which means that the black hole is thermodynamically stable.  Physically this is because its Hawking radiation is reflected off of the boundary and back into the black hole fast enough that the black hole reaches thermal equilibrium with its Hawking cloud and thus does not evaporate.  In the CFT description these stable black holes are the generic states of the CFT at high temperature, and in particular the scaling of their entropy with energy matches that of the dual CFT at high temperature:
\be
S\propto E^{\frac{d-2}{d-1}}.
\ee    
We can also ask about the CFT description of the full AdS-Schwarzschild wormhole: for sufficiently high temperature this is given by the ``thermofield double state'' 
\be
|TFD\ran=\frac{1}{\sqrt{Z}}\sum_i e^{-\beta E_i/2}|i^*\ran_L|i\ran_R,
\ee
where $|i\ran_R$ are the set of energy eigenstates of the CFT on a spatial sphere, $|i^*\ran_L$ are the energy eigenstates of the CFT on another copy the spatial sphere (the $*$ indicates that these are related by a certain antiunitary operator we won't discuss), and $E_i$ are the energy eigenvalues.  There are two copies of the CFT because there are two asymptotic boundaries, as in figure \ref{penroseadsschfig}.  This state is called the thermofield double because the reduced state on either copy is just the thermal density matrix with inverse temperature $\beta$.  To address the information problem, we now consider the thermal two point function 
\be
C(t)\equiv \frac{1}{Z}\Tr(e^{-\beta H} O(t) O(0)),
\ee
where $O(t)$ is a real scalar primary operator in the CFT which is dual to some real scalar bulk field as in \eqref{extdict} (we've suppressed its position-dependence).  From the bulk point of view, a simple calculation using quantum field theory in the AdS-Schwarzschild background shows that this correlator decays exponentially at late times.  This can be viewed as a form of information loss, roughly speaking since the information about a perturbation at $t=0$ becomes arbitrarily small at late times.    On the other hand if a black hole is really a quantum system with a finite entropy given by the Bekenstein-Hawking formula \eqref{SBH}, then the energy eigenvalues should have a discrete spectrum and such exponential decay should be impossible.  In fact given a plausible assumption called the ``eigenstate thermalization hypothesis'' we can estimate the long time average of $C(t)$ to be (see \cite{Harlow:2014yka} for more details)
\be
\lim_{T\to \infty}\int_0^Tdt |C(t)|^2\approx e^{-2S},
\ee
so the exponential decay must eventually end. This is a version of the information problem: bulk effective field theory suggests that $C(t)$ should go to zero, while unitarity suggests that it should bottom out at values of order $e^{-S}$.  In AdS/CFT we know that the latter is correct: up to some potential subtleties involving large strings or membranes near the AdS boundary, the energy spectrum of any CFT with a semiclassical dual is discrete, with level spacings of order $e^{-S}$, and in such a theory $C(t)$ must at late times on average be of order $e^{-S}$. 

\bfig
\includegraphics[height=5cm]{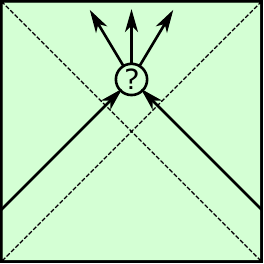}
\caption{A scattering experiment behind the horizon of the AdS-Schwarzschild geometry.}\label{bhscatterfig}
\efig
We've now seen that we can learn quite a bit about quantum gravity using the extrapolate dictionary \eqref{extdict}.  On the other hand  it is not the case that every question of possible interest about the gravity theory can be studied in this way.  A particularly simple example which can't is shown in figure \ref{bhscatterfig}: using the dictionary \eqref{extdict} we can prepare an initial state involving two particles which enter the wormhole from opposite sides and then scatter.  We cannot however probe the final state of this scattering by using \eqref{extdict}: the particles which are produced end up in the future singularity, and thus there is no obvious way to study them using properties of the dual CFT.  If we wish to know what kinds of particles were produced in this scattering, we need to ``back off'' the $r\to \infty$ limit in \eqref{extdict} and learn how to represent operators which are located deep in the bulk directly in the dual CFT, a problem which is called \textit{bulk reconstruction}.  

\bfig
\includegraphics[height=4cm]{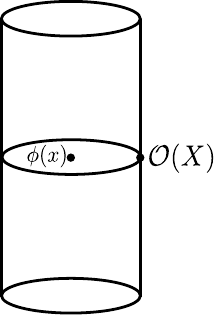}
\caption{The commutator paradox of AdS/CFT: how can a local operator in the center of the bulk commute with all boundary local operators on a time slice and yet be a nontrivial boundary operator?}\label{bulkcommfig}
\efig
There are various approaches to bulk reconstruction, but an essential observation is that any CFT representation of a bulk field away from the boundary will necessarily have a limited regime of validity, and even on states where it is valid it will only be an approximate notion.  One way to see this is via something called the \textit{commutator paradox} \cite{Almheiri:2014lwa}.  Let $x$ be a point in the middle of a bulk Cauchy slice $\Sigma$ and $X$ a boundary point on the boundary $\partial\Sigma$ of this Cauchy slice (see figure \ref{bulkcommfig}).  Since $x$ and $X$ are spacelike separated in the bulk, locality in the radial direction suggests that we should have
\be\label{bulkcomm}
[\phi(x),O(X)]=0
\ee
for any bulk field $\phi$ and local CFT operator $O$.  On the other hand it is a basic feature of quantum field theory, in formal circles referred to as the ``time-slice axiom'', that the set of local operators on a time slice should generate the full set of operators on the Hilbert space.  This means that any operator which computes with all local boundary operators $O(X)$ on a time slice must be proportional to the identity.  Applying this axiom to the CFT representation of $\phi(x)$, from \eqref{bulkcomm}  we apparently reach the conclusion that $\phi(x)$ must be proportional to the identity, which is certainly not a property we expect of any nontrivial local field!   The way out of this paradox is to realize that in most of the states in the dual CFT have a bulk description consisting of a big black hole (with $r_s\gg \ell_{ads}$) which has swallowed the point $x$, and when $x$ is far inside such a black hole it is not so clear that it still needs to obey \eqref{bulkcomm}.  The above contradiction with the time-slice axiom is removed once we only require \eqref{bulkcomm} to hold in a subspace of states where no such black hole is present.  On the other hand \eqref{bulkcomm} tells us that this subspace of states must have rather remarkable properties, since the information about what is going on in the center of the bulk must be present in the boundary degrees of freedom but not be accessible by any local measurement.  This however is not impossible, and in fact it is the hallmark of \textit{quantum error correction}, which is a set of protocols which were developed to protect quantum computers from environmental noise.  This idea can be developed in considerable detail, showing that AdS/CFT indeed provides a remarkable example of a quantum error correcting code, and this observation clarifies many aspects of the correspondence \cite{Almheiri:2014lwa}.  

\bfig
\includegraphics[height=5cm]{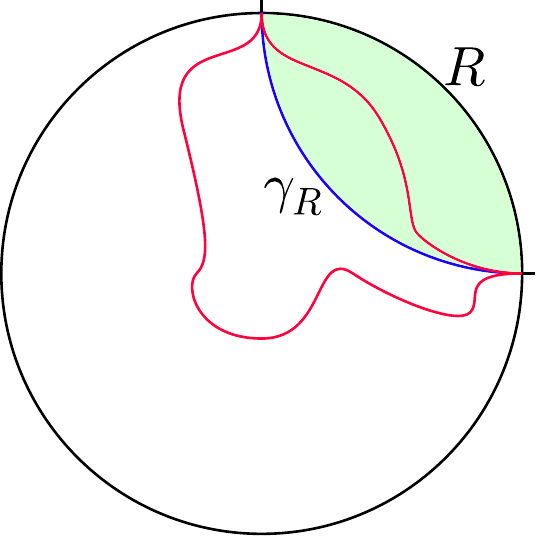}
\caption{The Ryu-Takayangi formula for $AdS_3/CFT_2$.  The red lines are non-minimal surfaces, while the Ryu-Takayanagi surface $\gamma_R$ is shaded blue.  Since space is two-dimensional here, $\gamma_R$ is a geodesic.  The homology region $H$ is shaded green.}\label{RTfig}
\efig
There is an additional piece of the AdS/CFT dictionary which has an especially close relationship both to black hole information and quantum error correction: the \textit{quantum extremal surface formula}.  This gives a bulk formula for the von Neumann entropy of any spatial subregion in the boundary CFT.  The formula has gone through several iterations, with the first version being proposed by Ryu and Takayanagi in 2006 \cite{Ryu:2006bv}:
\bi
\item \textbf{Ryu-Takayanagi (RT) formula:} Let $\rho$ be a $CFT_d$ state which is dual to an approximately classical bulk geometry that has time-reversal symmetry, and let $\Sigma$ be a bulk Cauchy slice which is time reversal invariant.  Moreover let $R\subset \partial \Sigma$ be a spatial region of the boundary CFT.  Then the von Neumann entropy on $R$ is given by
\be
S(\rho_R)=\min_\gamma \frac{\mathrm{Area}(\gamma)}{4G},
\ee
where $\gamma$ varies over all $(d-1)$-dimensional surfaces in $\Sigma$ which are \textit{homologous} to $R$, meaning there exists a $d$-dimensional \textit{homology hypersurface} $H\subset \Sigma$ such that $\partial H = R\cup \gamma$.  The minimal surface, which we'll call $\gamma_R$, is often called the \textit{RT surface} (or an RT surface if it isn't unique).
\ei
We illustrate the basic idea in figure \ref{RTfig}.  This formula contains the Bekenstein-Hawking formula as a special case, since if we take $\rho$ to be the thermofield double state and $R$ to just be one of the two boundaries then this von Neumann entropy is precisely the thermal entropy and $\gamma_R$ becomes precisely the horizon (see figure \ref{RT2fig}).  It is much more general however, as it can compute the von Neumann entropy also for regions were the reduced state is not thermal. 
\bfig
\includegraphics[height=4cm]{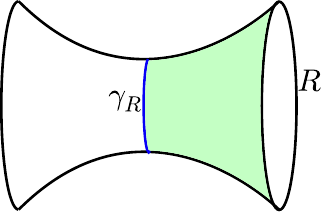}
\caption{The RT surface $\gamma_R$ for the right half of the thermofield double.}\label{RT2fig}
\efig

The RT formula has passed many interesting checks, and can even be in some sense derived, but it also has several obvious shortcomings.  In particular the restriction to states with time-reversal invariance is rather arbitrary, and it neglects quantum effects.  These defects were gradually fixed by various authors in a sequence of papers, resulting in the modern version of the formula \cite{Engelhardt:2014gca}: 
\bi
\item \textbf{Quantum Extremal Surface (QES) formula:} Let $\rho$ be any state of a $CFT_d$ which has a semiclassical bulk description, and $R$ be any spatial subregion of the boundary CFT.  Then the von Neumann entropy of $\rho$ on $R$ is given by 
\be\label{qes}
S(\rho_R)=\min_\gamma\left(\underset{\gamma}{\mathrm{ext}}\left(\frac{\mathrm{Area}(\gamma)+\ldots}{4G}+S_{bulk}(\rho_H)\right)\right),
\ee
where $\gamma$ ranges over $(d-1)$-dimensional spatial surfaces in the bulk which are \textit{homologous} to $R$ in the sense that there exists a bulk spatial region $H$ such that $\partial H=R\cup \gamma$.  The ``$\ldots$'' indicate higher order local terms integrated over $\gamma$, which must be determined from the effective action (which we assume still is dominated by ``Einstein plus matter'' in the infrared).  The surface $\gamma_R$ that wins is called the \textit{quantum extremal surface}.  
\ei
In this definition ``semiclassical'' means there is a semiclassical expansion in $G$ around some state of definite geometry, or perhaps superposition of an $O(1)$ number of such geometries.  This proposal has various appealing features:
\bi
\item The quantity $\frac{\mathrm{Area}(\gamma)+\ldots}{4G}+S_{bulk}(\rho_H)$ is what Bekenstein called the ``generalized entropy'' $S_{gen}$, and it has been shown to obey a second law in many situations and also to be UV-finite and universal even though each term separately is not.  
\item The ``bulk entropy term'' clearly needs to be there in some form, as we could simply create a local excitation near the boundary the middle of the region $R$ which carries some $O(1)$ entropy and this must somehow be reflected in $S(\rho_R)$ even though it doesn't affect the area.
\item In simple situations $S_{bulk}$ will be $O(1)$, in which case if there is time-reversal symmetry this reduces to the RT proposal.  
\ei
Moreover it has been shown that any quantum error correcting code naturally obeys a version of the quantum extremal surface formula, and so \eqref{qes} can be understood as a consequence of the quantum error-correcting nature of the AdS/CFT correspondence \cite{Harlow:2016vwg}.  

\bfig
\includegraphics[height=5cm]{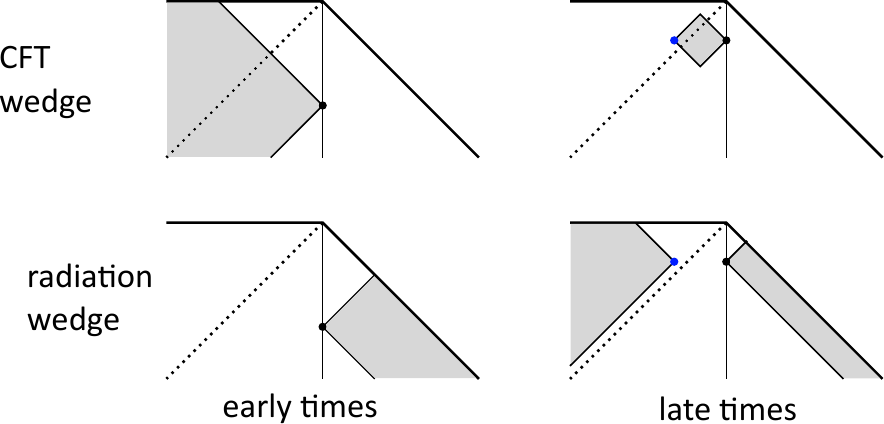}
\caption{Entanglement wedges for the calculation of the Page curve.  To the left of the vertical line is the gravitating region, while to the right is the non-gravitating bath.  The black hole horizon is the dotted line.  At early times the entanglement wedge of CFT includes most of the black hole interior, while at late times it doesn't.  At early times the entanglement wedge of the radiation includes just the radation, while at late times it also includes an ``island'' which is bounded by the new quantum extremal surface found in \cite{Penington:2019npb,Almheiri:2019psf}, which is represented as the blue dot.}\label{pageqesfig}
\efig
In the last few years it has been realized that the quantum extremal surface formula can be directly used to verify that the Page curve of certain evaporating black holes is consistent with unitarity \cite{Penington:2019npb,Almheiri:2019psf}.  The details of these calculations go beyond the scope of this article, but I'll give a brief sketch of how this works and what it means.  The basic idea is the following:
\bi
\item[(1)] Begin with a big black hole in AdS space, meaning a black hole with $r_s\gg \ell_{ads}$, which is dual to a highly energetic thermalized state in the CFT description.
\item[(2)] Locally couple the CFT to the boundary of a ``reservoir'' system, consisting of a non-gravitational field theory living on half of a Minkowski space in one dimension higher than the dual CFT.  From the bulk point of view this adds an asymptotically flat region to the spacetime, into which the black hole can radiate.  
\item[(3)] Evolve the system forward, causing the black hole to gradually evaporate into the reservoir.
\item[(4)] Use the quantum extremal surface formula to compute the entropy of the CFT and the reservoir as a function of time.
\ei
This quantum extremal surface calculation is illustrated in figure \ref{pageqesfig}. The key insight is that at late times a new quantum extremal surface appears, shown as a blue dot in figure \ref{pageqesfig}, whose generalized entropy eventually is less than that of the naive quantum extremal surface (which is given by the empty set).   This new surface is closed to the horizon, so its generalized entropy is essentially given by the horizon area in Planck units.  This is decreasing with time, giving the curve shown as the solid line in figure \ref{pageoptionsfig}.  There are several mysterious aspects of this calculation, with perhaps the most mysterious being why applying the QES formula to Hawking's ``wrong'' picture of the dynamics gives the ``right'' entropy.  The answer to this question, and more generally the explanation for what is really going on, seems to be that the black hole interior is encoded into the fundamental degrees of freedom via a ``non-isometric code protected by computational complexity'': there are states of exponential computational complexity in the effective field theory description which are annihilated by the holographic map, but in a way that does not disrupt Hawking's picture of the bulk for simple observables \cite{Akers:2022qdl}.  Many details of this picture remain to be worked out, but there is a marked sense in the community that a real resolution of Hawking's paradox is at last taking shape.  
  
\section{Conclusion}
We have now covered quite a lot of ground, and it may be of value to quickly review the main points.  
\bi
\item Quantum fields on either side of a black hole horizon are entangled, and this causes black holes to radiate at a temperature $T_{Hawking}=\frac{1}{8\pi GM}$.  Over time this energy loss causes black holes to evaporate, leading to an apparent contradiction with the unitarity of quantum mechanics.  Resolving this tension requires a major modification of the semiclassical structure of spacetime: spacetime itself must be emergent.
\item To understand how to resolve this puzzle we can turn to string theory, which is our best theory of quantum gravity so far.  In situations with a sufficient amount of supersymmetry string theory allows us to explicitly count the microstates of black holes, in all cases finding results which are consistent with the semiclassical analysis of Bekenstein and Hawking.  
\item String theory also leads to the discovery of the AdS/CFT correspondence, which says that quantum gravity in asymptotically-AdS spacetimes is equivalent to conformal field theory living at the asymptotic boundary.  As the latter is manifestly unitary, this gives a unitary theory of black hole physics.  And indeed the ``bulk'' spacetime is emergent: it only arises in certain states and in some approximation.  The mathematics of this emergence is quantum error correction.
\item In recent years we have understood that AdS/CFT correspondence with sufficient precision to allow for explicit calculations of the radiation entropy as a function of time, also known as the Page curve, for certain evaporating black holes.  These calculations gives results which are consistent with unitarity.  
\ei
I have attempted to give an appetizer for all of these ideas.  There are of course many details which have been suppressed, and the references are a natural first place to look for some of them.  The field continues to progress rapidly, including many developments which I have not had space to discuss, and it is likely to remain exciting for years to come.  

\paragraph{Note:} This chapter is the pre-print of the version currently in production. Please cite this chapter as the following: 

D.Harlow. “Black holes in quantum gravity” in The Encyclopedia of Cosmology (Set 2): Black Holes, edited by Z. Haiman (World Scientific, New Jersey, 2023).
\bibliographystyle{jhep}
\bibliography{bibliography}
\end{document}